\documentclass{rstransa}

\jname{rsta}
\Journal{Phil. Trans. R. Soc.}

\begin{document}
\title{Tracking the national and regional COVID-19 epidemic status in the UK using weighted Principal Component Analysis}
\author{Ben Swallow$^{1}$, Wen Xiang$^{2}$, Jasmina Panovska-Griffiths$^{3,4}$}
\address{$^{1}$School of Mathematics and Statistics, University of Glasgow, Glasgow, UK\\
$^{2}$Department of Statistics, London School of Economics and Poltical Science, London, UK\\
$^{3}$The Big Data Institute, Nuffield Department of Medicine, University of Oxford, Oxford, UK\\
$^{4}$The Queen's College, University of Oxford, Oxford, UK}

\subject{Statistics, Bioinformatics, Computational Biology}

%%%% Keyword entries to be placed here %%%%

%%%% Insert corresponding author and its email address}
\corres{Ben Swallow\\
\email{ben.swallow@glasgow.ac.uk}}

\begin{abstract}
[This paper has now been accepted in the journal under DOI 10.1098/rsta.2021-0302.]

One of the difficulties in monitoring an ongoing pandemic is deciding on the metric that best describes its status when multiple intercorrelated measurements are available. Having a single measure, such as the effective reproduction number R, has been a simple and useful metric for tracking the epidemic and for imposing policy interventions to curb the increase when R >1. While R is easy to interpret in a fully susceptible population, it is more difficult to interpret for a population with heterogeneous prior immunity, e.g., from vaccination and prior infection. We propose an additional metric for tracking the UK epidemic which can capture the different spatial scales. These are the principal scores (PCs) from a weighted Principal Component Analysis. In this paper, we have used the methodology across the four UK nations and across the first two epidemic waves (January 2020-March 2021) to show that first principal score across nations and epidemic waves is a representative indicator of the state of the pandemic and are correlated with the trend in R. Hospitalisations are shown to be consistently representative, however, the precise dominant indicator, i.e. the principal loading(s) of the analysis, can vary geographically and across epidemic waves.

\end{abstract}
\begin{fmtext}
\end{fmtext}

\maketitle

\section{Introduction}\label{introduction}

Throughout the SARS-CoV-2 pandemic, there has been consistent requirements from modellers to 'nowcast' the pandemic i.e  provide a quantitative assessment of the current state of the pandemic and to 'forecast' the epidemic i.e. predict the future trajectory at varying spatial and temporal scales. There are inherent difficulties in modelling the current state of a pandemic, with challenges associated with data censoring, reporting lags, uncertainty in cause of death, asymptomatic cases and testing errors all contributing to possible biases in informed measures of incidence (Shadbolt et al., 2021; Swallow et al., 2022). While several metrics can be used to provide a quantitative evaluation of the current state of an epidemic, the effective reproduction number, R, has been the most commonly and widely used metric (see Pellis et al. (2021) for further discussion of this). R is a measure of the number of secondary infections stemming from a single infection and reflects the transmisibility or infectiousness of the viral variant during an epidemic. It allows tracking of the status of the epidemic with $R<1$ suggesting that the epidemic is in the declining phase, whereas $R>1$ describes increased transmission and a growing epidemic. In the UK it is derived as a consensus range from a number of different models and reported weekly by the UK Health Security Agency (UKHSA (2020)). Since the onset of the pandemic in the UK, this R consensus range has been used to track the epidemic status and to inform and guide policy decision makers in imposing and removing interventions such as imposing reduced social interactions intervention i.e. lockdowns. However, while R is a useful measure, it is sensitive to, for example, the choice of data being used (e.g., the number of tests being carried out and any delays in reporting of cases) or the method for calculating it (e.g., the combination and type of models used or the length of the time-slice being used for the calculation). It can also be computationally demanding to provide accurate estimates. Furthermore, once the population is partially or fully vaccinated defining population-wide R may not suffice. There are also difficulties associated with the fact that the onset of symptoms often happen a few days after the onset of infection, and hence the current value of R really represents the state of the pandemic at some point in the recent past; in the UK the consensus is that the current value of R is lagged by 2 to 3 weeks (UKHSA (2020)). 

Other metrics such as the rate of daily hospital admissions have been suggested as alternative metrics that can be used alongside R, especially in a very heterogeneous population with mixed immunity from vaccination or from prior infection with a specific viral variant. In this paper we develop and apply a statistical method to derive a different metric related to the daily cases, hospitalisations, mechanical ventilation bed (MVB) admissions or deaths related to COVID-19 to track the national and regional epidemic status. For this purpose, we analyse the time series of these metrics to determine the important temporal, spatial and mechanistic dimensions of the data and highlight potential outliers in the time series. 

We utilise multivariate projection methods, specifically methods of dimension reduction, aiming to find lower-dimensional representations of multiple (correlated) measurements to provide new measurement axes that are weighted as a linear combinations of the original measurements. Principal Components Analysis (PCA) is one such method which also has the added advantage of representing a geometric rotation of the data into orthogonal, or mutually independent, axes that are maximal in terms of variance retention. Given there would \emph{a priori} be expected to be a high degree of correlation between the time series measuring varying dimensions of the pandemic status, dimension reduction techniques would be expected to project the data matrix down into a much smaller number of uncorrelated bases. Larger numbers of infected cases generally lead to higher numbers of those requiring medical intervention or mortality rates. However, within a horizon of emerging variants which may be more transmissible but possible less severe, this relationship may not be linear nor consistent across measurements and settings.

There appears to have been little attempt to account for these inherent correlations between measurements thus far. The formal structures inherent in existing models are not always directly applicable to simple statistical approaches, and can often be computationally complex when conducting inference. Methods that enable practitioners to obtain indicators of the current or recent state of the pandemic in a quick and efficient manner are therefore highly desirable.

We aim to do this by developing an approach similar to Xiang and Swallow (2021) in analysing data from the UK COVID-19 dashboard (``Coronavirus (COVID-19) in the UK,'' 2021) to study simpler representations of the multivariate output of cases, deaths, hospitalisations and MV bed occupancy. We use S-Mode and T-mode PCA, with temporal weight matrix calculated from median correlation in residuals following a Generalised Additive Model fitted to a smooth of date index. We conduct the PCA on deaths, cases, hospitalisations and MV beds both at a UK level, and separately for the four UK nations. We also conduct additional analyses looking for differences between the dynamics of the first two principal COVID-19 waves. The analyses aim to explore whether a single epidemic metric or a combination of metrics can be a useful indicator of the status of the epidemic across nations and waves, and hence be potentially useful to track in future alongside R and growth rate. 

{%
\section{Methods}\label{methods}}

{%
\subsection{Principal Components Analysis}\label{principal-components-analysis}}

Multivariate projection and decomposition methods, such as Principal Components Analysis (PCA), enable the extraction of structures in multivariate data through an eigen-decomposition of the correlation or covariance matrix. The eigenvectors form a rotated basis of the data, or equivalently a new set of uncorrelated axes that are ordered by magnitude of their corresponding eigenvalues. This corresponds directly to the proportion of variation in the original measurements that they explain. 

For a data matrix \(\mathbf{X}\) of dimension $n\times p$, a principal components analysis can be conducted through a singular value decomposition (SVD) of the column mean-centred matrix. The SVD decomposes \(\mathbf{X}\) into 

\begin{equation*}
    \mathbf{X}=\mathbf{U}\mathbf{D}\mathbf{V}^{\intercal},
\end{equation*}

\noindent where \(\mathbf{U}\) and \(\mathbf{V}\) are the column matrices of left and right singular vectors respectively and \(\mathbf{D}\) a diagonal matrix of singular values. The objective is to find a linear transformation
\(\mathbf{Y} = \mathbf{U}^{\intercal}\mathbf{X}\), where
\(\mathbf{U}^{\intercal}=(U_{11},...,U_{p1})^{\intercal}\) is a matrix of constants such that the
\(\mathbb{V}ar(\mathbf{Y}_1)\) is maximised, subject to the normalising
constraint \(\mathbf{U}^{\intercal}\mathbf{U}=\mathbf{I}\). It can therefore be seen both as a variance-maximisation projection of the covariance matrix or a linear transformation into an orthogonal set of bases. As the singular values are ordered by magnitude, or hence proportion of variation explained, for highly colinear data, lower order approximations to can be produced by setting these singular values to zero.

Extensions to standard PCA relax the \emph{a priori} assumption of unknown structure and allow users to account for existing spatial and/or temporal structures inherent in the data through the use of spatial and/or temporal weighting matrices (Baldwin et al. 2009). Accounting for existent temporal structures in the data allow the extraction of important residual joint structures that can be more readily interpreted than if these known structures are not accounted for. In conventional PCA it is common practice to standardise the data matrix by column standard deviations. Standardising by a corresponding weight matrix ensures that the temporal structures are more comparable and are not dominated by measurements that are in larger units or are leading lags.  

Alternative rotations, often referred to as S-Mode or T-Mode PCA of the data can lead to different bases. Assuming the rows of the matrix correspond to the time points of the data and the columns the individual time series, the S-Mode PCA aims to find dominant temporal trends across the four data streams. Conversely, T-Mode PCA is conducted on the transpose of the matrix and aims to find different patterns across the time series and the associated time points at which they occur.

{%
\subsection{Flow directed PCA}\label{flow-directed-pca}}

{%
\subsubsection{Weighted PCA}\label{temporal-weight-matrix}}

Adjusting PCA to account for known spatial and/or temporal structure in the data can support the extraction of novel trends in the data, as well as account for spatial variability in units across space or temporal lags in the time-series. Assuming the data matrix is a \(n \times p\) matrix, a \(p \times p\) column weight matrix \(\boldsymbol\Omega\) and \(n \times n\) row weight matrix \(\boldsymbol\Phi\) can be constructed so that PCA is applied to a transformed matrix \(\boldsymbol{\tilde{X}}= \boldsymbol\Phi\boldsymbol{X}\boldsymbol\Omega\). 

In our analyses here, we only consider the temporal column matrix, \(\boldsymbol\Omega\), as the spatial aspect is less evident when considering measurements of different patient status, however we describe the full process for completeness. Weighting or supervising the PCA by a temporal matrix should transform the data matrix towards a  standard multivariate normal, so that the subsequent PCA selects eigenvectors with a high ratio of spatial or inter-measurement variability relative to temporal variation. This reduces the potential of over-fitting by prioritising variation between measurements, whilst still ordering the eigenvalues by maximal variance retention. 

Let \(\boldsymbol{\tilde{X}}\) denote the scaled data matrix, weighted by spatial and/or temporal weight matrices as follows.

\[\boldsymbol{\tilde{X}}= \boldsymbol\Phi\boldsymbol{X}\boldsymbol\Omega=\boldsymbol{\tilde{U}\tilde{D}\tilde{V^{\intercal}}},\]

where the decomposed matrices generated on the scaled variables are denoted with a tilde. In this instance, $\Phi$ is the spatial weight matrix and $\Omega$ the temporal weight matrix. As only a temporal matrix is used here we remove dependence on $\Phi$, concentrating only on the temporal matrix $\Omega$. Hence, the PCs of the new weighted variables become \(\boldsymbol{X\Omega\tilde{V}}\), and the loadings are \(\boldsymbol{\Omega^{-\intercal}\tilde{V}}\).

The temporal weight matrix is constructed similarly to Gallacher et al. (2017) using independent Generalised Additive Models (GAMs) (Wood, 2017). For measurement $Y$ we fit the model

\begin{equation*}
    \mathbb{E}[Y]=g^{-1}\Big(\beta_0 + \sum_{k=1}^{K}f_k(x)\Big),
\end{equation*}

\noindent where the $f_k(.)$ are smooth functions, often represented as splines and $g()$ is the link function mapping to the scale of the response. These models are fitted by restricted maximum likelihood (REML) in the mgcv package (Wood, 2012) to each of the time series with an intercept and an univariate smooth function of date as predictor, that is a univariate $x$. As in Xiang and Swallow (2021), this aims to remove trends specific to each stream, with residual variation used to determine the principal scores. Remaining correlation in the model residuals between time \([1,\dots, (n-1)]\) and \([2,\dots, n]\) is calculated for each stream and then the median value is used as an estimate of \(\rho\), the average global temporal correlation. The median residual temporal correlations \(\rho\) are estimated for each of the output time series (these will vary in each analysis depending on the spatial resolution/measurements used). The \(i\)th row and \(j\)th column element of the temporal weight matrix \(T_{i,j}\) is then specified as \(\rho^{|i-j|}\) for all time indices in the original data matrix. The weight matrix $\Omega$ is then taken as the matrix square-root of $T$, that is $\Omega=T^{\frac{1}{2}}$. The temporal weight matrix is applied to the data matrix and dimension reduction is then conducted to create a new uncorrelated set of bases. The method is identifiable up to a change in sign, so in some cases similar trends are apparent, only inverted.

\subsubsection{S-Mode and T-Mode Principal Components Analysis}\label{fd-PCA}

Next we describe how these methods can be applied to explore important global spatial and temporal trends in cases and deaths from COVID-19. To extract the important trends, PCA and similar dimension reduction techniques are an obvious choice. PCA conducts an eigen-decomposition of the covariance (or correlation) of a data matrix, with eigenvalues ordered by magnitude to reduce a set of \(p\) correlated variables to a smaller set of \(k<p\) orthogonal variables. Versions of PCA for spatio-temporal data were referred to by Richman (1986) as S-mode and T-mode PCA, the particular mode depending on whether the columns of are time points (T-mode) or time series index (S-mode).

S-Mode PCA aims to find dominant temporal trends across the spatial locations, highlighting a small number of dominant temporal trends across all countries and/or time series. Conversely, T-Mode PCA aims to find different spatial patterns in the data and the associated time points at which they occur. In general, however, PCA finds unsupervised structures in the data by conducting an eigen-decomposition of the correlation of covariance matrix. Whilst this can often be useful in visualising data in lower dimensions, it is not possible to guide the structure of the new axes using prior information or independent data. In order to account for known spatio-temporal correlations inherent in the data, we use spatio-temporally weighted S-mode and T-mode PCA, which aim to find dominant temporal and spatial patterns respectively. Gallacher et al. (2017) extended these approaches to account for known spatio-temporal structures in river flow systems through the use of spatial and/or temporal weight matrices to inform spatio-temporal structure. Analyses were conducted using the stpca package in R.

S-Mode PCA will be conducted with the columns of matrix in each formulation being the the particular data stream, and the rows corresponding to time points (day). For T-Mode PCA, the columns correspond to the time points and the rows are the data streams. In S-Mode, the output of interest will be a series of time-indexed points projected into the dimensions of principal variance, denoted the principal scores. In T-mode, the scores will be a projection of the time points into a single score for each of the data streams. In both cases, the principal scores will be a linear combination of the original higher-dimensional data with loadings representing the contribution of each of the original measurements to the corresponding score. It would be expected these scores are centred around zero, with approximately symmetric variation around this. Any deviations from this would suggest outliers or patterns that warrant further consideration.

{%
\subsection{Data}\label{data}}

Data used in this study are extracted from the UK COVID-19 Dashboard (``Coronavirus (COVID-19) in the UK,'' 2021) and consist of daily measurements of reported cases, deaths, hospitalisations and MV beds occupied (representing hospital occupancy where ventilation support was required) between 2020-04-02 to 2021-02-22 (327 days of 16 measurements). This time period spans the first two waves of the pandemic in the UK, and terminates at the point that the vaccination programme was rolled out on a wide scale. Numbers were available at the level of the four individual UK nations, as well as aggregated across the UK as a whole. The nationally segregated data were used for all analyses, with some analyses run both jointly across all nations and independently for each nation, to determine differences in results depending on spatial scale. Data were checked for inconsistencies and outliers that may have impacted results, although none were found.

For additional analyses, the data were also stratified into nation-specific data matrices and matrices corresponding to the two individual waves, namely March to May 2020 (59 days) representing the first epidemic wave and September 2020 and April 2021 (175 days) representing the second epidemic wave (as per ONS, 2021), were also constructed. All matrices are column mean-centred prior to analysis to ensure easier comparison between dimensions.

The data and computer code used to run the analyses and generate figures reported in this paper are openly available at \url{https://doi.org/10.5281/zenodo.6078749}. 

{%
\section{Results}\label{results}}

\hypertarget{pooled-uk-analysis}{%
\subsection{Combined UK analysis}\label{pooled-uk-analysis}}

\begin{figure}
\centering
\includegraphics[width=0.7\textwidth]{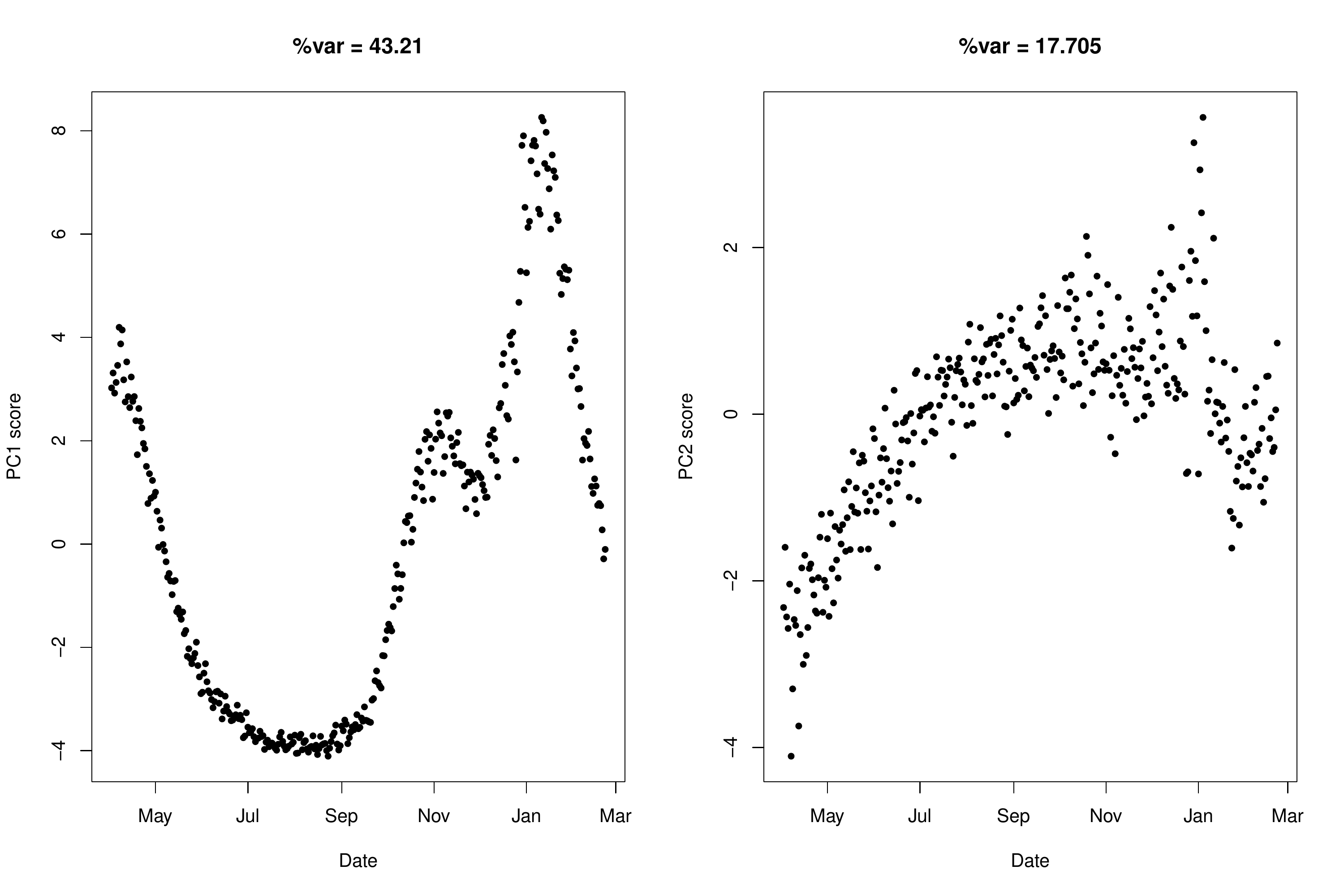}

\caption{\label{fig:smodall}First 2 PC scores for temporally weighted S-Mode PCA for pooled UK data, with corresponding proportion of variance explained above each. Data correspond to April 2020 to March 2021.}
\end{figure}

As a result of the PCA analysis we derived the first and second principal scores (PCs). The initial results were obtained at the UK-wide scale suggest that a single combined index of the four different measurements from each of the four nations is able to explain approximately 42\% of variation, with the second PC accounting for a further 18\% (Figure \ref{fig:smodall}), with loadings being roughly equal. The observed trend in PC1 is incredibly smooth, with an initial peak very early in the studied period before a steep decline in the state of the pandemic towards the summer months. The overall state then worsened from early September through to early November, when a lockdown was implemented. The release of restrictions over the Christmas period from late December leads to a drastic increase in the overall observed trend until early January, when further interventions were introduced as well as the initialisation of the vaccination roll-out. Following this point, there is a steep linear reduction in the observed overall state to the end of the modelled period in Mar. 

In addition, there is a further trend observed in the second PC, which shows a general increase in pandemic state from April through to September, after which it remains relatively stable. Given the relatively smooth nature of this second score and its asymptotic behaviour, this would suggest a more subtle and smooth change in the observed behaviour. The change in testing capacity over this time-frame increased significantly, particularly up until the autumn/winter period. Spearman correlation coefficient was calculated as 0.36 between the second score and reported number of new tests (``Coronavirus (COVID-19) in the UK'', 2021), giving some support to the idea this second score could be associated with an underlying change in the data collection, rather than necessarily a change in the pandemic.

\begin{figure}
\centering
\includegraphics[width=\textwidth]{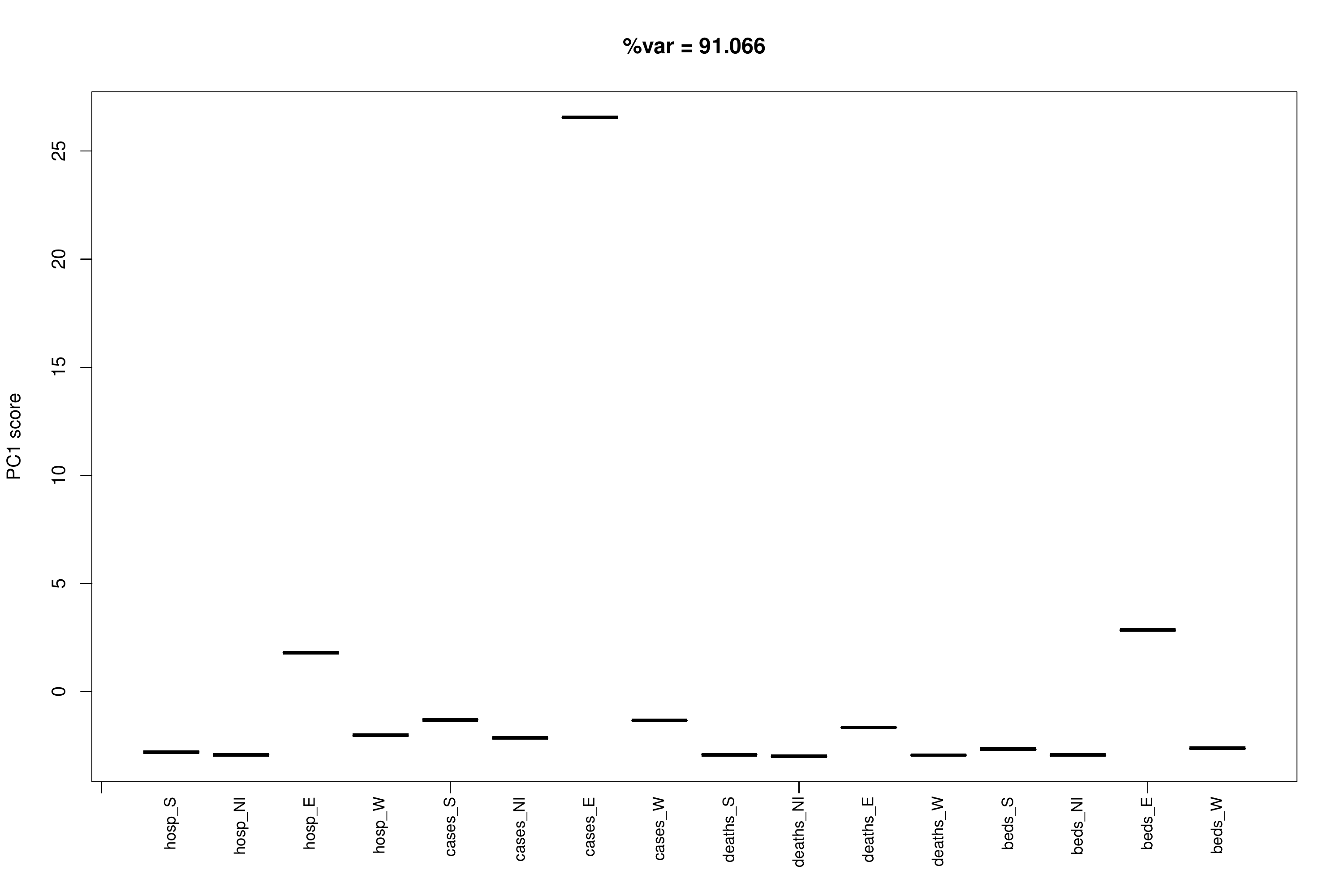}
\caption{\label{fig:tmodallu}First PC score for temporally-weighted T-Mode PCA in a combined UK analysis. The approach highlights measures that deviate markedly from the others/average.}
\end{figure}

Figure \ref{fig:tmodallu} shows the first PC score for the T-mode analysis. The aim of this approach is to detect time series that deviate significantly from each other. In this analysis, only a single PC is required to explain over 90\% of the variation in the data, suggesting further PCs are not informative. Following temporal standardisation, it would be expected that these follow approximately a multivariate normal distribution. Most of the scores lie around zero, with some small deviations for hospitalisations and MV beds in England. What is particularly noteworthy, however, is the significant deviation for cases away from the central trend, suggesting that case numbers in England, and to a lesser extend hospitalisation and MV beds, deviated markedly away from trends in the other nations and other time series.

\begin{figure}
\centering
\includegraphics[width=\textwidth]{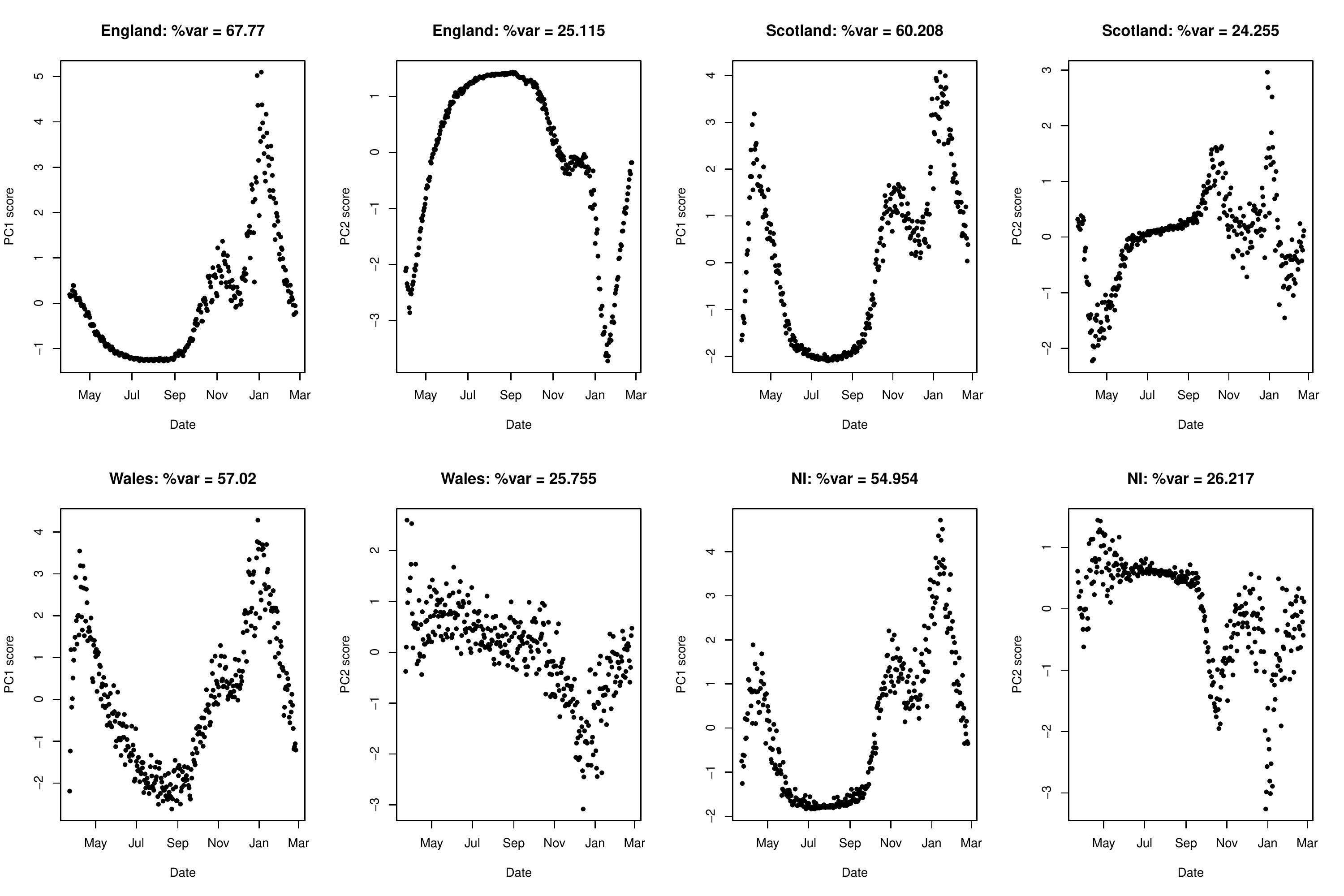}
\caption{\label{fig:smodN}First 2 PC scores for temporally weighted S-Mode PCA across the different UK nations.}
\end{figure}

{%
\subsection{Nation-specific analysis}\label{nation-specific-analysis}}

\begin{figure}[!h]
\centering
\includegraphics[height=0.7\textheight]{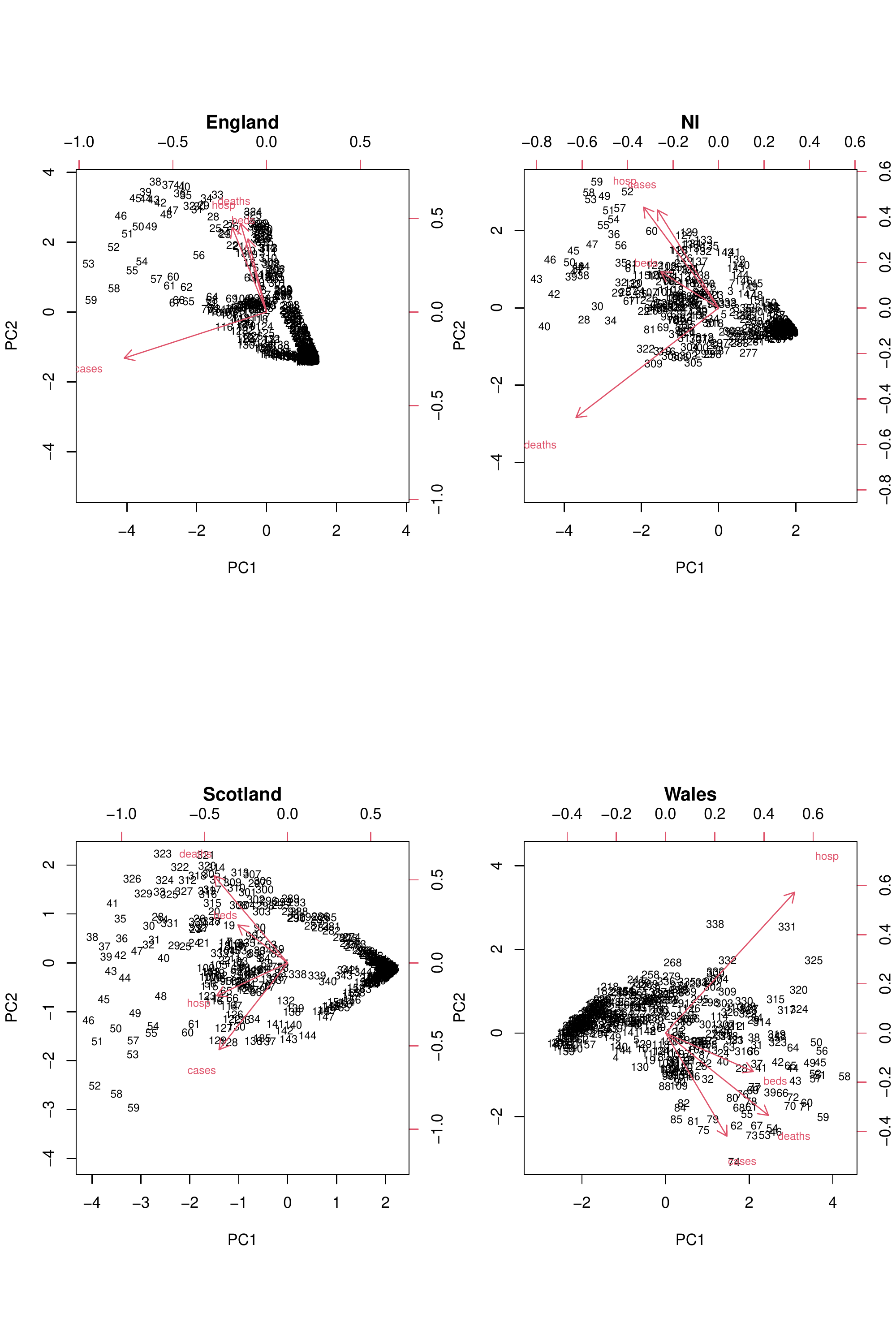}
\caption{\label{fig:smodNload}Biplots for the temporally weighted S-Mode PCA across the different UK nations. Length of the arrows corresponds to the loading contribution of that measure to each of the first two PCs.}
\end{figure}

For the nation-specific analyses, the first S-mode PC was able to explain between 65\% and 70\% of the total variation in the data, with around 25\% explained by PC2. This suggests that there is a single dominant trend, with a second less-dominant independent trajectory occurring (Figure \ref{fig:smodN}). The corresponding loadings show that the dominant contributor to the principal axis varies across nations (Figure \ref{fig:smodNload}). For England, it was cases; Northern Ireland deaths; Scotland showed roughly equal contributions; and Wales hospitalisations.

All analyses showed peaks in April 2020 and January 2021, with a smaller peak in November 2021 that showed initial signs of reduction before increasing again to the major peak in January 2021. The lockdown that was introduced in November 2020 seems to have been particularly beneficial in Scotland, which showed a reduction down to similar low levels as over the summer months in 2020. The subsequent significant growth up to January 2021 follows temporary Christmas reductions in distancing measures and may be a result of this. A combined impact of strict intervention measures after the Christmas holidays and in the New Year, combined with the rolling out of the vaccination programme for susceptible individuals, lead to a reduction across all trends and this is highlighted in the first PC score. 

Whilst Scotland, Wales and Northern Ireland show a short sharp peak in April 2020 that drops sharply down to September, the first PC score in England showed a much shallower change over the initial months in 2020 before similar dynamics over the winter 2020 to the other nations. The change in the early months observed in the other three nations is relegated to PC2 in England, suggesting that overall changes were less dramatic than in other areas of the UK. This pattern in PC2 for England relative to the other nations, suggesting the earlier wave was distinctly different from the first wave.

\begin{figure}
\centering
\includegraphics[width=\textwidth]{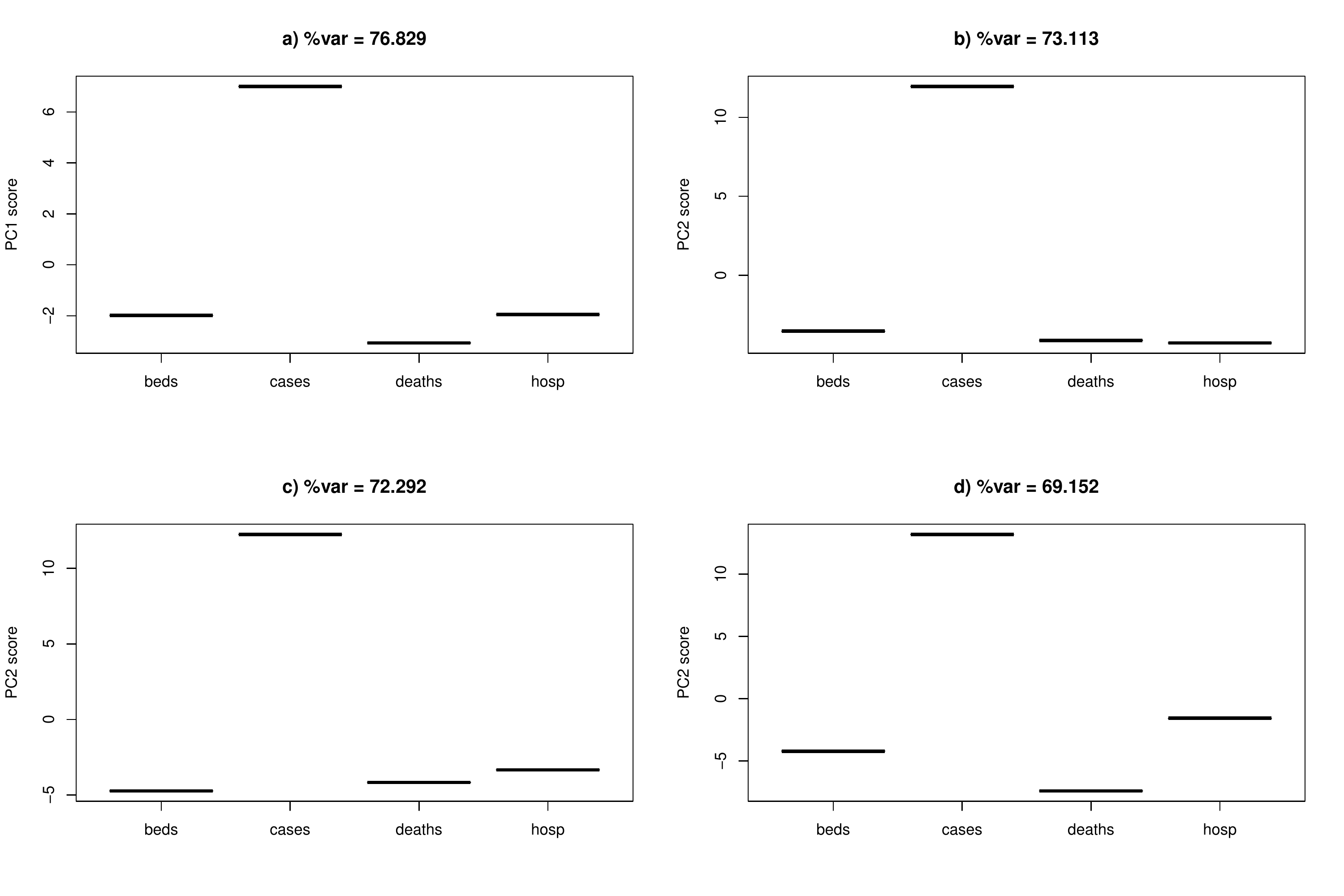}
\caption{\label{fig:tmodeacht}First PC score for temporally weighted T-Mode PCA - a) England; b) Northern Ireland; c) Scotland; d) Wales.}
\end{figure}

The T-mode analysis by nation (Figure \ref{fig:tmodeacht}) suggests that hospitalisations are consistently close to zero when projected into the new axes. Cases vary markedly from this, suggesting that they are very different and therefore caution should be taken when considering them as representative of the epidemic state.

{%
\subsection{Comparison of waves}\label{comparison-of-waves}}

\begin{figure}
\centering
\includegraphics[width=0.7\textwidth]{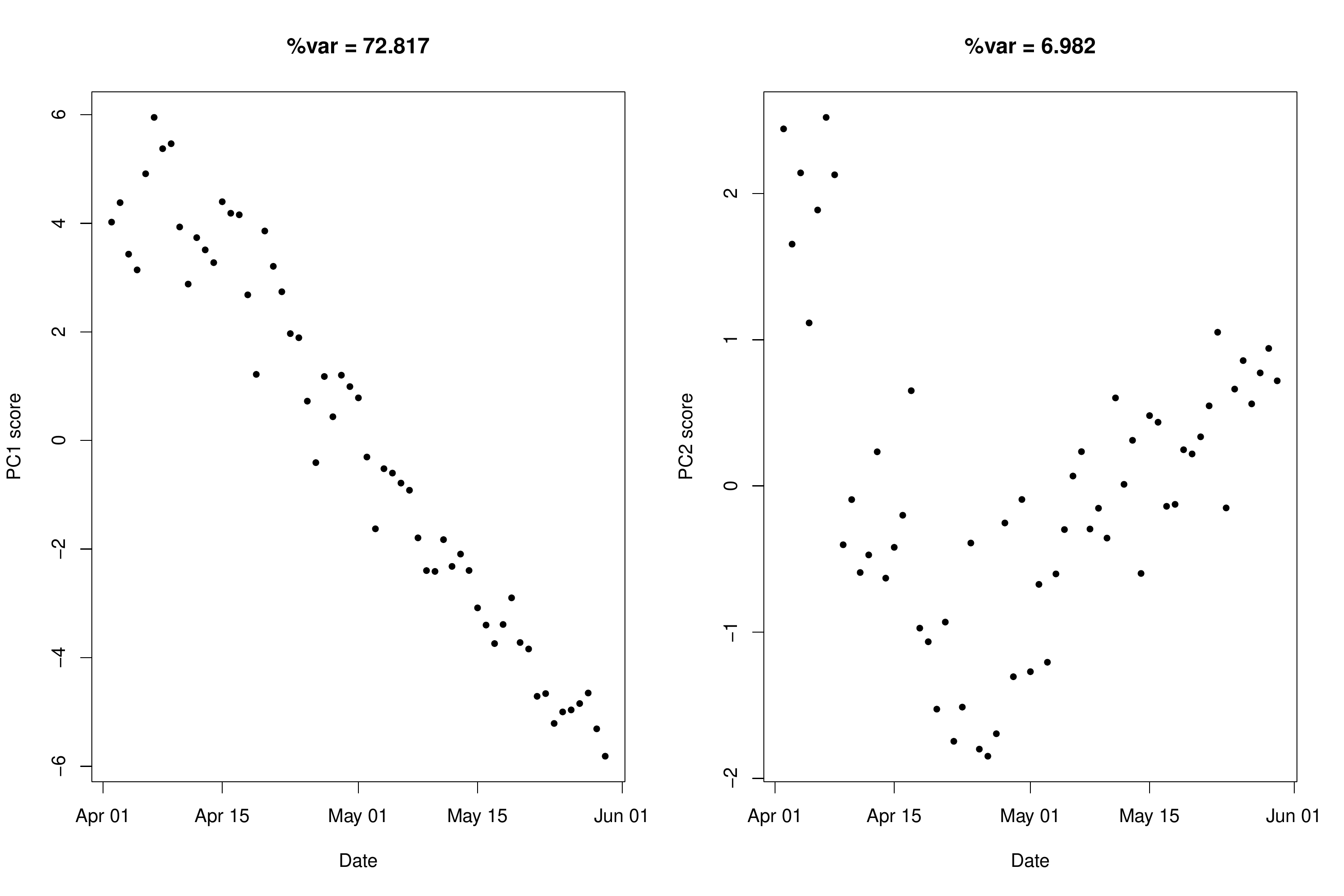}
\caption{\label{fig:W1}First 2 PC scores for temporally weighted S-Mode PCA covering dates in COVID-19 Wave 1, April to June 2020.}
\end{figure}

\begin{figure}
\centering
\includegraphics[width=.7\textwidth]{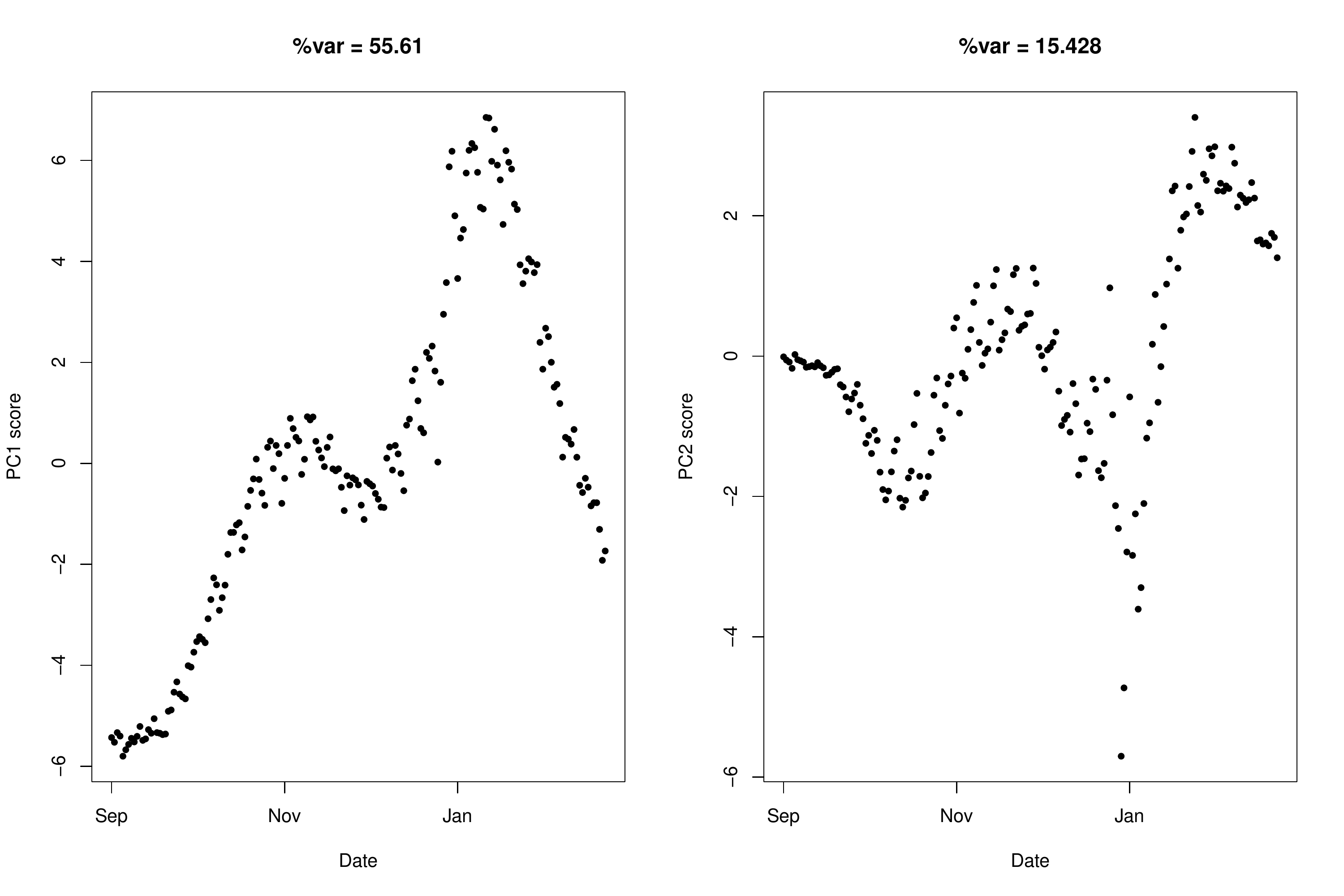}
\caption{\label{fig:W2}First 2 PC scores for temporally weighted S-Mode PCA covering dates in COVID-19 Wave 2 spanning September 2020 to March 2021.}
\end{figure}

We also rerun the unified S-Mode analysis to compare dynamics of trends separately for the two principal waves of COVID-19 over the studied period (Figures \ref{fig:W1} and \ref{fig:W2}). We denote wave 1 to consist of data between 20th January 2020 and 31st May 2020 and wave 2 to correspond to data from 1st September 2020 to the end of March 2021 (ONS, 2021). The results were consistent with the full analysis in terms of which variables dominated the first principal component, however there were clearly different general temporal trends in each of these periods. Particularly noteworthy is the principal linearly increasing trend seen in PC1 for wave 1 dynamics. PC2 for this analysis explains little of the variance, so is not worth focusing on. Wave 2 shows more variation through time, including an interrupted linear growth which temporarily decreases at the time of a national lockdown in November. A similar linear reduction is seen at the end of wave 2. Both of these time periods correspond to strict lockdown periods in the UK.

The loadings between the two analyses, which represent the contribution of an individual measurement to each linear transformation, showed some interesting trends. Specifically, in wave 2, the loadings decreased for hospitalisations in all nations; deaths in England; and cases in Wales and Northern Ireland. All other measurements were more dominant in wave 1.

\subsection{Relationship with $R$}

\begin{figure}
\centering
\includegraphics[height=.5\textheight]{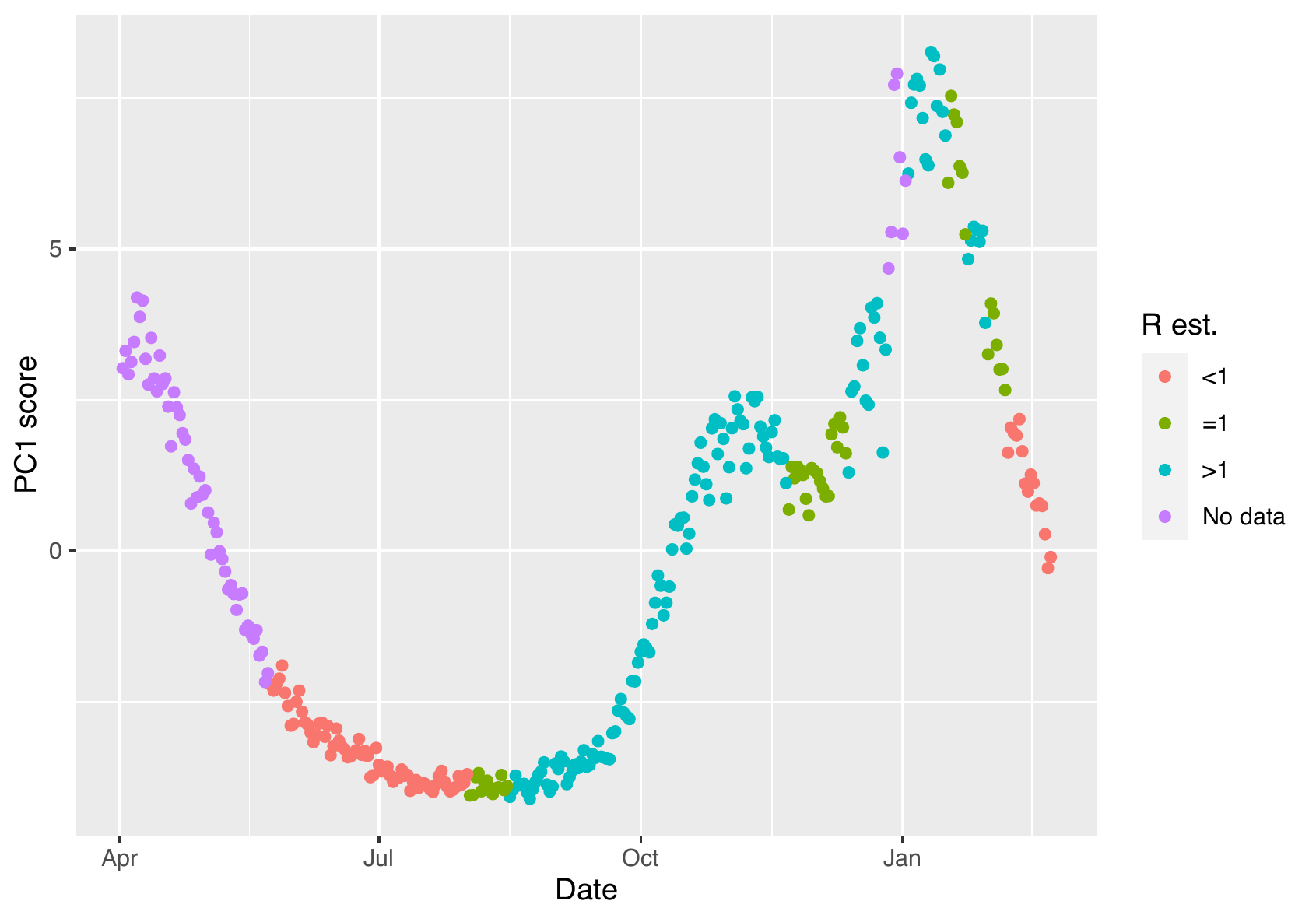}
\caption{\label{fig:R0}First PC score from Figure \ref{fig:smodall} coloured by strata of the upper bound of the estimated UK reproductive number upper bound from (UKHSA 2020) in the same week. Red is R<1; green is R=1 and teal is R>1. Purple corresponds to weeks in which no estimate was provided.}
\end{figure}

\begin{figure}
\centering
\includegraphics[height=.5\textheight]{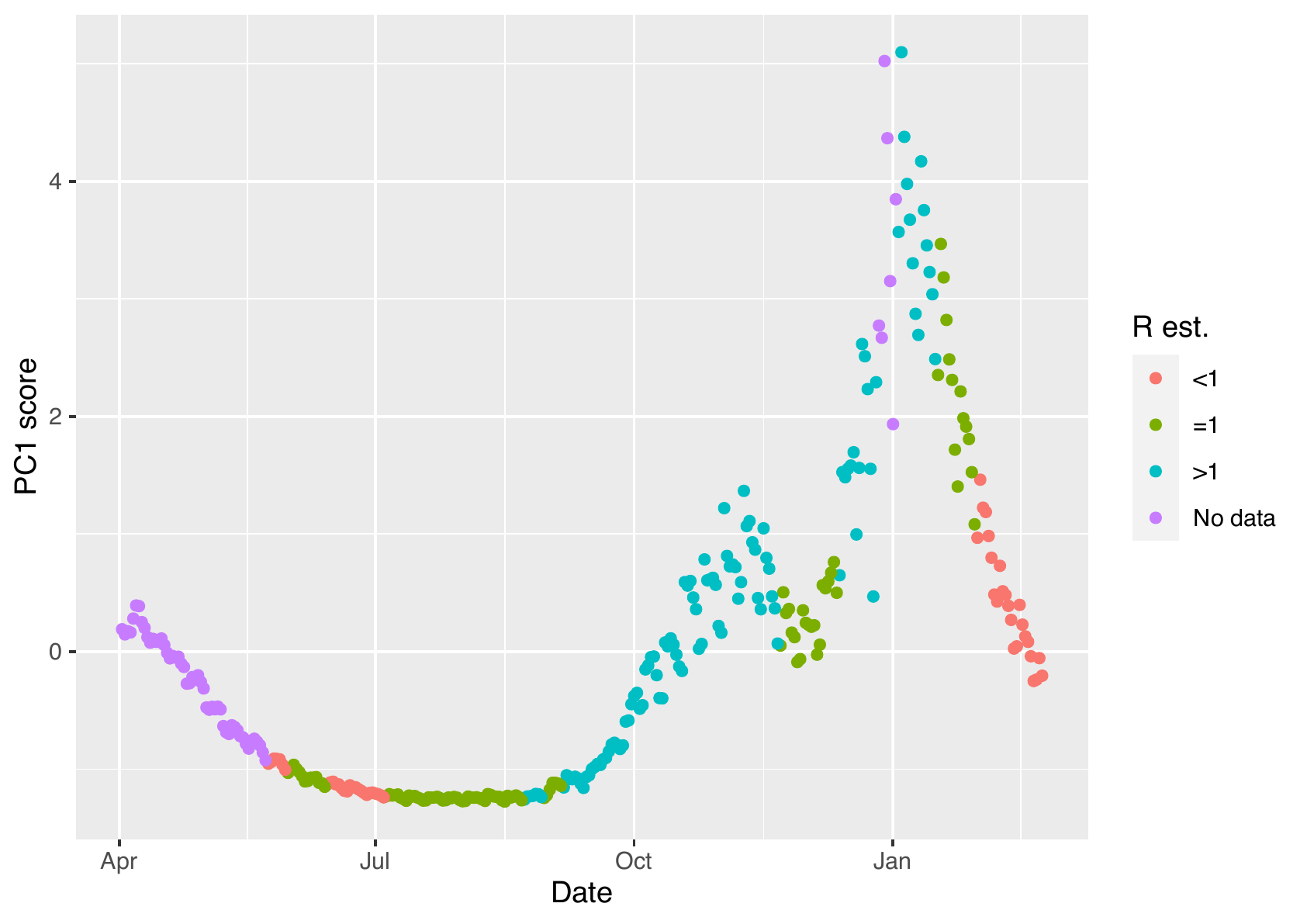}
\caption{\label{fig:R0E}First PC score from S-Mode PCA in England coloured by strata of the upper bound of the estimated UK reproductive number upper bound from (UKHSA, 2020) in the same week. Red is R<1; green is R=1 and teal is R>1. Purple corresponds to weeks in which no estimate was provided.}
\end{figure}

Finally we compare the results form the full S-mode analysis in Figure \ref{fig:smodall} and the England-specific analysis with the corresponding estimates of UK-wide reproductive number from (UKHSA 2020). Comparable data from the other nations was not publicly available. Both upper and lower bounds of the reproduction number R were available, generated from an ensemble of epidemiological models with the upper bound being used in Figure \ref{fig:smodall}. During periods of increase R, there is very good agreement in the trend between the time-series of R and the corresponding observed increase in the first PC score (Figures \ref{fig:R0} and \ref{fig:R0E}). Similarly, when R drops and remains below one, we also observe a inflection in the first PC score. This agreement suggests that the principal PC score is strongly correlated with R (Spearman's coefficient $\rho=0.84$ and $\rho=0.72$ for UK and England respectively) and hence principle PC may feasibly provide a viable additional quantity for fast nowcasting of the pandemic.

\hypertarget{conclusions}{%
\section{Conclusions}\label{conclusions}}

Overall across nations, our results suggest that hospitalisations are consistently important as a representative indicator of the state of the pandemic at any given time point, being highlighted as outside the standard linear combinations of the other variables in most analyses. This matches the fact that these are likely to be relatively unbiased compared to cases (Vekaria et al., 2021) and less age- or health status-specific than MV beds. MVBs are also expected to be limited by capacity unlike the overall hospital admissions, the former of which policy has changed significantly over the course of the pandemic (NICE (2021)). The combination of results from S-mode and T-mode analyses at different scales allow for a greater understanding of the pandemic at different levels.

In some analyses, the epidemic trends can also be driven by the number of reported cases, notably in Scotland where non-pharmaceutical interventions (NPIs), particularly in the populated central belt incorporating Glasgow and Edinburgh, were maintained much longer over the study period so the dynamics would be expected to be different guided by these NPIs' impact. For example, Greater Glasgow and neighbouring authorities remained under varying levels of restrictions between early September 2020 and May 2021, which were particularly severe between November and December. However, reported numbers of cases can be highly variable and subject to possible under-reporting and strongly dependent on testing, contact-tracing and isolation behaviour and policies. 

For the nation-specific analyses, variation was observed in the dominant contributor to the principal axis, suggesting that simple averages of different measures are not appropriate. Numbers of hospitalised patients appear to be consistently highlighted as outliers in the data, however. This is aligned with hospitalisations and especially occupancy levels, being a trend that, together with R, has been important in driving policy decisions over the first and second epidemic waved i.e. over the study period here.

Deaths and MVBs as a proxy for hospital occupancy have also been used as epidemic indicators by policy decision makers. In our analysis, however, they don't appear to be a significant indicator of the epidemic status. We note that this could be due to possible bias in these as for example reported death figures have been challenging during the pandemic due to delays in reporting and uncertainty over exact cause of death. For instance, there have been several different figures adopted e.g., mortality within 28 days of a positive test, deaths with COVID-19 on the death certificate and deaths with rather than from COVID-19. All of these metrics represent deaths related to COVID-19, but all have aspects of the data can cause biases and give uncertainty in reporting deaths related to COVID-19. The number of MVBs, whilst perhaps slightly less biased than the reported numbers of cases and deaths, is likely to be susceptible to changes in policy and severely limited by a carrying capacity at each hospital and movement of patients in and out of these wards depending on need. Hospitalisation figures, indicating admission to hospital with COVID-19 and are less susceptible to either of these biases (Vekaria et al., 2021). Hence our suggestion that overall hospitalisations may be a good epidemic indicator and an additional potential metric to track alongside R and  growth rate is sensible.

The good agreement between our approach and independent estimates of the reproduction number in England gives support to our approach as a complementary fast, real-time method for determining the state of the pandemic from multivariate noisy data. As we had mentioned previously, tracking R in a highly heterogeneous population with an imprinted prior immunity from vaccination and prior infection with different variants is very different to R in a unvaccinated population or fully susceptible one at the onset of the epidemic or after immunity has fully waned. Furthermore, the current tracking of the epidemic status via the nowcasting process by the modelling team at the UKHSA is reliant on a combination of mathematical models that are calibrated to vast amount of currently available data from the UK COVID-19 dashboard (UKHSA, 2020). Looking towards a situation where less data may be produced daily e.g., when potentially only hospitalisations may be tracked rather than also cases and deaths related to COVID-19, an alternative method to assess epidemic status such as the one proposed here will be of great value.

Via our step-by-step statistical analysis we highlight the care that must be taken when choosing the level of aggregation of data for statistical analysis (Jeffery et al., 2009; 2014). For example, our results highlight that aggregating  the data to UK level changes the dynamics of the data streams and masks important local dynamics. 

There were limited differences in general trends between the first and second wave of COVID-19 across the UK, with both waves showing linear reductions post-lockdown interventions. MVBs generally became less dominant while hospitalisations emerged as more dominant in the second wave. Deaths in England were also more dominant in wave 1. This could relate to a change in policy only intervening in the most serious cases, or an improvement in outcome due to improved understanding of the disease and the impact from the large scale vaccination programme from December 2020.

The fact that most analyses of spatial trends show two major principal components, whilst temporal analyses tend to show only one dominant trend in space, is worth highlighting. The principal spatial trend will generally show the dominant trend over larger regions, whilst the second spatial trend could relate to an increase in testing capacity over the time period, showing an overall increase in numbers detected over the whole, whereas early in the emergence of SARS-CoV-2, only those with severe symptoms would likely be detected. It could also highlight second-order structure in the data, corresponding to higher variability in times of policy changes (e.g., lockdowns, reopening of schools, reopening of public services) or local variations in trends not picked up by the main principal component.

One of the drawbacks of conventional PCA is that as an unsupervised method, it is often challenging to interpret the often abstract outputs that it produces. Also in many scenarios, known structures in the data can be readily available and are not themselves of interest. One of the interests in these scenarios is finding latent structures in the data that explain hidden correlations between measured quantities. The study of these structures can then inform further on possible important mechanisms within the system under study.

In Xiang and Swallow (2021), the authors analysed data from global trends relating to reported cases of, and deaths resulting from, SARS-CoV-2. Their analyses found a single temporal trend dominated the global spread of the disease. This suggests that there are multi-scale processes occurring, namely a general spread across countries that happens smoothly and consistently, and more local dynamics within a country that occur at a finer resolution. This again highlights the importance of considering appropriate scales when looking at dynamics of infectious diseases (e.g., Garabed et al., 2020). Aggregation of data may be beneficial computationally or in terms of reducing the impact of biases and errors in the data, however it will inevitably change results.

Removing common temporal correlation from the data through a temporal weight matrix reduced the amount of variation explained in all analyses, but had little impact on the overall conclusions. This supports the idea that there is a great deal of similarity across nations within the UK in terms of seasonal dynamics and principal trends, however it is not sufficient to treat the streams as either independent or entirely analogous.

There appear to have been few attempts to look at the structures inherent in multiple measurements of the pandemic and how to reduce these down to single measures, other than the R value widely applied. Rahman et al. (2020) construct a structural equation model, in which dimension reduction is included, and a variable relating to `pandemic severity' is generated as one of their dimensions of interest. However, this is not the main focus of their research and they only include cases and deaths in their analysis. Their analyses support the idea that treating the measurements as independent is clearly erroneous, as they clearly have large amounts of correlation. Also, errors and biases in each stream will be compounded if independent models are fitted. Using dimension reduction techniques, such as directed PCA, allows for the correlation to be incorporated into the methodology and reduces the impact of any single stream bias by integrating over them all.

In summary, in this study we have used an established statistical technique to generate an alternative, yet complementary indicator of the epidemic status to R and growth rate. We have used the PCA methodology described here to show that overall, across nations and epidemic waves, the level of hospitalisations with COVID-19 a good indicator of the epidemic status. However, the precise best indicator, i.e. the principal score(s) of the PCA, vary geographically and across epidemic waves.   

\hypertarget{references}{%
\section{References}\label{references}}

\hypertarget{refs}{}
\leavevmode\hypertarget{ref-ONSdash}{}%
Baldwin, M. P., Stephenson, D. B., Jolliffe, I. T. (2009). Spatial Weighting and Iterative Projection Methods for EOFs, Journal of Climate, 22(2), 234-243. 
\url{https://doi.org/10.1175/2008JCLI2147.1}

\hypertarget{refs}{}
\leavevmode\hypertarget{ref-ONSdash}{}%
Coronavirus (COVID-19) in the UK {[}WWW Document{]}, 2021. {[}WWW Document{]}. URL \url{https://coronavirus.data.gov.uk/details/download}

\leavevmode\hypertarget{ref-gallacher17}{}%
Gallacher, K., Miller, C., Scott, E.M., Willows, R., Pope, L., Douglass, J., 2017. Flow-directed PCA for monitoring networks. Environmetrics 28, e2434. \url{https://doi.org/10.1002/env.2434}

\leavevmode\hypertarget{ref-Garabed}{}%
Garabed, R.B., Jolles, A., Garira, W., Lanzas, C., Gutierrez, J., Rempala, G., 2020. Multi-scale dynamics of infectious diseases. Interface Focus 10, 20190118. \url{https://doi.org/10.1098/rsfs.2019.0118}

\leavevmode\hypertarget{ref-Jeffery}{}%
Jeffery, C., Ozonoff, A., Pagano, M., 2014. The effect of spatial aggregation on performance when mapping a risk of disease. International Journal of Health Geographics 13, 9. \url{https://doi.org/10.1186/1476-072X-13-9}

\leavevmode\hypertarget{ref-Jeffery2009}{}%
Jeffery, C., Ozonoff, A., White, L.F., Nuño, M., Pagano, M., 2009. Power to detect spatial disturbances under different levels of geographic aggregation. Journal of the American Medical Informatics Association 16, 847--854. \url{https://doi.org/10.1197/jamia.M2788}

\leavevmode\hypertarget{ref-Jeffery2009}{}%
NICE, 2021. COVID-19 rapid guideline: managing COVID-19. \url{https://www.nice.org.uk/guidance/ng191}

\leavevmode\hypertarget{ref-ONSsurvey}{}%
ONS, 2021. Coronavirus (COVID-19) Infection Survey technical article: waves and lags of COVID-19 in England, June 2021.\url{https://www.ons.gov.uk/peoplepopulationandcommunity/healthandsocialcare/conditionsanddiseases/articles/coronaviruscovid19infectionsurveytechnicalarticle/wavesandlagsofcovid19inenglandjune2021}

\leavevmode\hypertarget{ref-PellisBirrel}{}%
Pellis, L., Birrell, P.J., Blake, J., Overton, C.E., Scarabel, F., Stage, H.B., Brooks-Pollock, E., Danon, L., Hall, I., House, T.A., Keeling, M., Read, J.M., JUNIPER consortium, 2021. Estimation of reproduction numbers in real time: Conceptual and statistical challenges.

\leavevmode\hypertarget{ref-Rahman}{}%
Rahman, M.M., Thill, J.-C., Paul, K.C., 2020. COVID-19 pandemic severity, lockdown regimes, and people's mobility: Early evidence from 88 countries. Sustainability 12. \url{https://doi.org/10.3390/su12219101}

\leavevmode\hypertarget{ref-richman86}{}%
Richman, M.B., 1986. Rotation of principal components. Journal of Climatology 6, 293--335. \url{https://doi.org/https://doi.org/10.1002/joc.3370060305}

\leavevmode\hypertarget{ref-shadbolt21}{}%
Shadbolt, N., Brett, A., Chen, M., Mario, G., McKendrick, I.J., Panovska-Griffiths, J., Pellis, L., Reeve, R., Swallow, B., 2021. The challenges of data in future pandemics. \url{https://www.newton.ac.uk/documents/preprints/}

\leavevmode\hypertarget{ref-swallow21}{}%
Swallow, B., Birrell, P., Blake, J., Burgman, M., Challenor, P., Coffeng, L.E., Dawid, P., Angelis, D.D., Goldstein, M., Hemming, V., Marion, G., McKinley, T., Overton, C., Panovska-Griffiths, J., Pellis, L., Prober, W., Shea, K., Villela, D., Vernon, I., 2022. Challenges in estimation, uncertainty quantification and elicitation for pandemic modelling. Epidemics (accepted for publication)

\leavevmode\hypertarget{UKHSA2020}{}%
UKHSA, 2020. The R value and growth rate. \url{https://www.gov.uk/guidance/the-r-value-and-growth-rate}

\leavevmode\hypertarget{UKHSA2021}{}%
UKHSA, 2020. Reproduction number (R) and growth rate: methodology. \url{https://www.gov.uk/government/publications/reproduction-number-r-and-growth-rate-methodology}

\leavevmode\hypertarget{ref-Vekaria}{}%
Vekaria, B., Overton, C., Wiśniowski, A., Ahmad, S., Aparicio-Castro, A., Curran-Sebastian, J., Eddleston, J., Hanley, N.A., House, T., Kim, J., Olsen, W., Pampaka, M., Pellis, L., Ruiz, D.P., Schofield, J., Shryane, N., Elliot, M.J., 2021. Hospital length of stay for covid-19 patients: Data-driven methods for forward planning. BMC Infectious Diseases 21, 700. \url{https://doi.org/10.1186/s12879-021-06371-6}

\leavevmode\hypertarget{ref-woodGAM}{}%
Wood, S.N., 2017. Generalized additive models: An introduction with R, second edition, Chapman \& Hall/CRC texts in statistical science. CRC Press.

\leavevmode\hypertarget{ref-woodmgcv}{}%
Wood, S.N., 2012. {Mgcv: Mixed GAM computation vehicle with GCV/AIC/REML} smoothness estimation.

\leavevmode\hypertarget{ref-Xiang21}{}%
Xiang, W., Swallow, B., 2021. Multivariate spatio-temporal analysis of the global covid-19 pandemic. medRxiv. \url{https://doi.org/10.1101/2021.02.08.21251339}

{%
\section{Acknowledgements}\label{acknowledgements}}

BS is a member of the Scottish COVID-19 Response Consortium, which was undertaken in part as a contribution to the Rapid Assistance in Modelling the Pandemic (RAMP) initiative, coordinated by the Royal Society.

JPG's work was supported by funding from the UK Health Security Agency and the UK Department of Health and Social Care. This funder had no role in the study design, data analysis, data interpretation, or writing of the report. The views expressed in this article are those of the authors and not necessarily those of the UK Health Security Agency or the UK Department of Health and Social Care.

\end{document}